\documentclass[preprintnumbers,amsmath,amssymb,twocolumn]{revtex4}
\usepackage[colorlinks=true,citecolor=red,filecolor=green,linkcolor=blue,pdfnewwindow=true]{hyperref}
\usepackage{hyperref}
\usepackage{titlesec}
\titleformat{\section}
	{\center\bfseries}{\makebox[30pt][l]{\thesection}}{0pt}{}
\titleformat{\subsection}
	{\flushleft\bfseries}{\makebox[30pt][l]{\thesubsection}}{0pt}{}
\titleformat{\subsubsection}
	{\flushleft\itshape}{\makebox[30pt][l]{\thesubsubsection}}{0pt}{}
\titlespacing\subsection{0pt}{12pt plus 4pt minus 2pt}{0pt plus 2pt minus 2pt}
\titlespacing\subsubsection{0pt}{12pt plus 4pt minus 2pt}{0pt plus 2pt minus 2pt}
\usepackage[font=small]{caption}
\usepackage{subcaption}
\usepackage{color}
\usepackage{graphicx}
\usepackage{makeidx}
\usepackage{bm}
\usepackage[title]{appendix} 
\usepackage{tabu}
\usepackage{float}
\usepackage{chemarr}
\usepackage{url}

\begin{document}

\title{Lower bound of entropy production at short time scales for noise-driven stochastic systems}
\author{Mairembam Kelvin Singh$^1$}
\email{kelvin.phd.phy@manipuruniv.ac.in}
\author{R.K. Brojen Singh$^2$}
\email{brojen@jnu.ac.in}
\author{Moirangthem Shubhakanta Singh$^1$}
\email{mshubhakanta@yahoo.com}
\affiliation{$^1$Department of Physics, Manipur University, Canchipur-795003, Imphal, Manipur, India.\\ $^2$School of Computational $\&$ Integrative Sciences, Jawaharlal Nehru University, New Delhi-110067, India.}

\begin{abstract}
{\noindent}The second law of thermodynamics governs that nonequilibrium systems evolve towards states of higher entropy over time. However, it does not specify the rate of this evolution and the role of fluctuations that impact the system's dynamics. Entropy production quantifies how far a system is driven away from equilibrium and provides a measure of irreversibility. In stochastic systems, entropy production becomes essential for understanding the approach to nonequilibrium states. While macroscopic observations provide valuable insights, they often overlook the local behaviors of the system, governed by fluctuations. In this study, we focus on measuring the lower bound of entropy production at short time scales for generalized stochastic systems by calculating the Kullback-Leibler divergence (KLD) between the probability density functions of forward and backward trajectories. By analysing the entropy production across sliding time scales, we uncover patterns that reveal distinctions between local, small-scale dynamics and the global, macroscopic behavior, offering deeper insights into the system's departure from equilibrium. We also analysed the effects of switching to different types of noise or fluctuations and found that the observations at larger time scales provide no distinction between the different forms of noise while at short time scales, the distinction is significant.\\ 

\noindent\textbf{Keywords:} Entropy production; Kullback-Leibler divergence; Short time scales; Noise; Stochastic systems.
\end{abstract}
\maketitle

\section{Introduction}
In its simplest notion, entropy is a measure of the disorder, uncertainty, or randomness of a system. However, this simple definition is often considered misleading \cite{Marty}. Strictly speaking, entropy measures the amount of energy irretrievably dissipated into the surroundings. During an irreversible process in a system, there may be exchange of particles, energy, or both, with the surroundings. Entropy production (EP) represents the minimum amount of entropy generated by a system in a nonequilibrium state. According to the second law of thermodynamics, the entropy of a system always increases, implying that EP is always positive for nonequilibrium processes. However, this law does not quantify the degree of divergence from equilibrium \cite{Dewar}, \cite{Chiri}, \cite{Cocconi}, making the estimation of EP essential, albeit complex.

The estimation of EP has been an active area of research. Machine learning techniques have been developed to estimate EP directly from time series data \cite{Otsubo1}, \cite{Otsubo2}. The thermodynamic uncertainty relation (TUR) provides a lower bound for EP by analyzing average fluctuating probability currents \cite{Manikandan}, \cite{Van}. EP estimation is also closely tied to measuring system irreversibility by capturing time-reversal asymmetry and broken detailed balance \cite{Maes}.

Theoretical, experimental, and simulation-based approaches for EP estimation have been widely explored. Common case studies include colloidal particles \cite{Speck}, molecular motors and flashing ratchets \cite{Kapustin}. Such systems often exhibit stochastic trajectories, leading to inconsistencies in EP estimation as not all trajectories are identical. Therefore, many studies focus on single stochastic trajectories to ensure consistency \cite{Seifert}.

The Kullback-Leibler divergence (KLD) \cite{Kullback} is another statistical measure extensively used to estimate the lower bound of EP. It quantifies the divergence between a given distribution and a reference distribution. By comparing the distribution of forward trajectories with that of backward trajectories under the same conditions, KLD provides a robust and consistent measure of EP \cite{Roldan}, \cite{Parrondo}. Unlike trajectory-specific methods, KLD maintains consistency across simulations with varied trajectories, as its qualitative depiction of EP remains stable.

In this study, we determined KLD to calculate the lower bound of EP at short time scales using sliding time intervals of specific lengths. This approach reveals the system's local dynamics at short time scales, which often contrast with the ordered behavior observed at macroscopic scales. Additionally, we examined the effect of incorporating different forms of noise in the dynamics, moving beyond the commonly used Gaussian white noise, which lacks temporal correlations. It is found that the observations at larger time scales provide no distinction between the different forms of noise while at short time scales, the distinction is significant.

Section 2 outlines the methods employed, beginning with a generalized linear stochastic differential equation (SDE) solved using the Milstein method. The determination of KLD from the time series data and the construction of probability density functions are also illustrated. Section 3 presents the results and discussion. Finally, the paper concludes with an analysis of the implications and future prospects of our study.\\

\section{Methodology}
We consider a generalized linear SDE of the form \cite{Kloeden}:
\begin{eqnarray}
\label{SDE}
\dot{x}=a(x,t)+b(x,t)\zeta
\end{eqnarray}
where $a(x,t)$ is the drift term, $b(x,t)$ is the diffusion term, $\zeta$ represents the noise or fluctuations and $\dot{x}$ denotes the first-order time derivative.

To solve eq.\ \eqref{SDE}, we employed the Milstein scheme, an enhancement of the Euler-Maruyama method with higher convergence order \cite{Kloeden}, \cite{Milstein}. The scheme is expressed as:
\begin{eqnarray}
\label{Milstein}
x_{n+1}=&x_n+a(x_n, t_n)\Delta t+b(x_n, t_n)\Delta\phi\nonumber\\
&+\frac{1}{2}b(x_n, t_n)b^\prime (x_n, t_n)(\Delta\phi^2-\Delta t)
\end{eqnarray}
where $\Delta\phi=\zeta\Delta t$ and $b^\prime (x_n, t_n)$ is the first-order spatial derivative of the diffusion term.

To calculate KLD and the lower bound of EP, we generated the density functions of the forward and backward trajectories, obtained using the Milstein scheme [eq.\ \eqref{Milstein}]. For this, Kernel Density Estimation (KDE) with a Gaussian kernel was applied, using the Normal Rule of Thumb (NROT) method for bandwidth selection and equally weighted samples \cite{Scott}, \cite{Silverman}.

Backward trajectories were obtained by reversing the order of iterations, starting at the simulation end-points and end-times of the forward trajectories with the same time step. This yielded two density functions, $\rho$ and $\bar{\rho}$, representing the forward and backward trajectories respectively. Using these functions, KLD can be determined as:
\begin{eqnarray}
\label{KLD}
\mathcal{D}(\rho || \bar{\rho})=\int dx~\rho\ln\left(\frac{\rho}{\bar{\rho}}\right)
\end{eqnarray}

If $\langle\Delta S_{\text{tot}}\rangle$ denotes the average entropy production of the whole system, then, the determined KLD value gives its lower bound \cite{Roldan}, \cite{Parrondo}, i.e., $\displaystyle\langle\Delta S_{\text{tot}}\rangle\ge k\mathcal{D}(\rho || \bar{\rho})$, where $k$ is the Boltzmann constant. Throughout our study, we kept $k=1$, so that the estimated lower bound of EP is dimensionless. KLD is strictly positive or zero (when $\rho=\bar{\rho}$) \cite{Kullback}. Thus, the time-reversal asymmetry and breaking of detailed balance is prominent if we obtain a non-zero value of KLD.

KLD was computed at short time scales using two approaches: increasing time intervals and sliding time intervals. In the former, we begin by determining KLD at a small time interval at the beginning of the total time interval and slowly increase the length of the interval by a fixed amount until it covers the total simulation time. In the latter, a fixed-length time window was slid across the simulation, capturing local dynamics. Observations depend on the interval length, representing the small scale of observation.

We also explored different noise forms, including Gaussian white noise, pink noise and Lévy noise. While Gaussian white noise lacks temporal correlation, pink noise exhibits long-range correlation with an autocorrelation function that exhibits power law decay behavior. Pink noise appears in phenomena like river flows, atmospheric temperatures \cite{Bunde}, etc. Lévy noise, associated with stochastic processes with stationary, independent increments, thus exhibiting discontinuous trajectories, is common in random walks and complex systems such as anomalous diffusion, turbulent flows, seismic activity and financial stock markets \cite{Eliazar}, \cite{Ibe}.\\

\begin{figure*}
\begin{center}
\includegraphics[height=5.0cm,width=15.0cm]{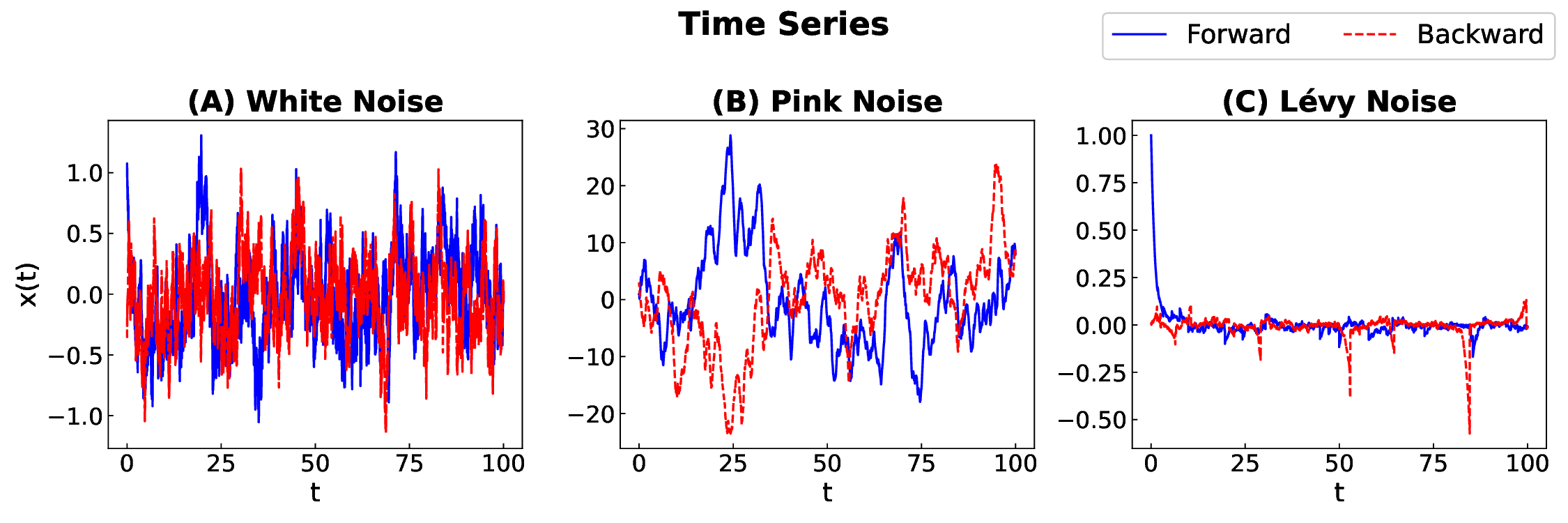}
\caption{\label{Fig1}Time series of eq.\ \eqref{SDE} with (A) Gaussian white noise (mean = 0, $\sigma=\sqrt{dt}$). (B) Pink noise ($\sigma=\sqrt{dt}$). (C) Lévy noise [mean = 0, $\alpha=1.5$, $\beta=0$ and scale = $\sigma(\Delta t)^{1/\alpha}$].}
\end{center}
\end{figure*}

\section{Results \& Discussion}
For all forms of noise, we used the initial condition, $x_0=1$, a total simulation time of 100 arbitrary units, and a step size, $\Delta t=0.001$. In the SDE given by eq.\ \eqref{SDE}, the drift and diffusion terms were set as $a(x,t)=-x$ and $b(x,t)=0.5$ respectively, for simplicity. To determine KLD at short time scales, we employed two approaches: increasing and sliding time intervals. For the increasing time intervals, we started with the interval 0-2 and increased the time window length, $W$ by 2 arbitrary units, progressing as 0-4, 0-6, and so on, until 0-100. For the sliding time intervals, we began with the interval 0-2 and slid it across the total simulation time in increments of 0.2 while maintaining the same time window length, $W$. This produced intervals like 0.2-2.2, 0.4-2.4, and so on, up to 98-100. We also observed the effect of using larger time window lengths of the sliding intervals. All measured quantities were dimensionless. When plotting KLD values, we used the endpoints of the increasing time intervals and the midpoints of the sliding time intervals.

For Gaussian white noise, we set the mean to 0. The standard deviation, $\sigma$ is set to $\sqrt{\Delta t}$ for both white noise and pink noise. Pink noise has a power spectral density inversely proportional to frequency. To generate it, we constructed a frequency-domain signal with uniformly distributed random phases, then applied an inverse real Fourier transform to revert back to time-domain. Lévy noise, characterized by heavy tails and power-law decay, requires additional parameters: the stability parameter, $\alpha$ (set to 1.5 for a tailed distribution) and skewness, $\beta$ (set to 0 for a symmetric distribution). The scale parameter of Lévy noise is normalized [scale = $\sigma(\Delta t)^{1/\alpha}$] to ensure that the generated noise is consistent with the time step, $\Delta t$ and the desired Lévy distribution properties. The Lévy noise samples were generated from a Lévy stable distribution that uses the Chambers-Mallows-Stuck (CMS) algorithm \cite{Chambers}, \cite{Nolan}.

The time series for white noise (Fig.\ \ref{Fig1}A) shows overlap between forward and backward trajectories at most points in time due to the lack of temporal correlation. However, for pink noise, forward and backward trajectories rarely overlap due to the long range temporal correlation (Fig.\ \ref{Fig1}B). The time series for Lévy noise (Fig.\ \ref{Fig1}C) shows typical sudden jumps. This asymmetry in the forward and backward trajectories will be reflected in the determination of KLD for short time scales.

\subsection{Increasing \& Sliding time intervals}
For both the approaches, we extracted the trajectory points for both the forward and backward trajectories and for each interval. A combined range of the two trajectories is maintained so that both the distributions are evaluated over the same domain. Using KDE, which is a non-parametric way of estimation of probability density function \cite{Scott, Silverman}, the density functions $\rho$ and $\bar{\rho}$ for the forward and backward trajectories are obtained. A small constant stability parameter, $\epsilon=10^{-8}$ is added to each density value to avoid computational error. After that the density values are normalised by their sum.

The density functions $\rho$ and $\bar{\rho}$ are used to determine the KLD values at each interval. As we obtained discrete values, the integration in eq.\ \eqref{KLD} is replaced by a summation performed over the discrete values of the particular interval. Also, due to the stochastic nature of the system, we determined the average KLD value at each interval, $\tilde{\mathcal{D}}$, for 30 simulations.

\begin{figure*}
\begin{center}
\includegraphics[height=5.0cm,width=15.0cm]{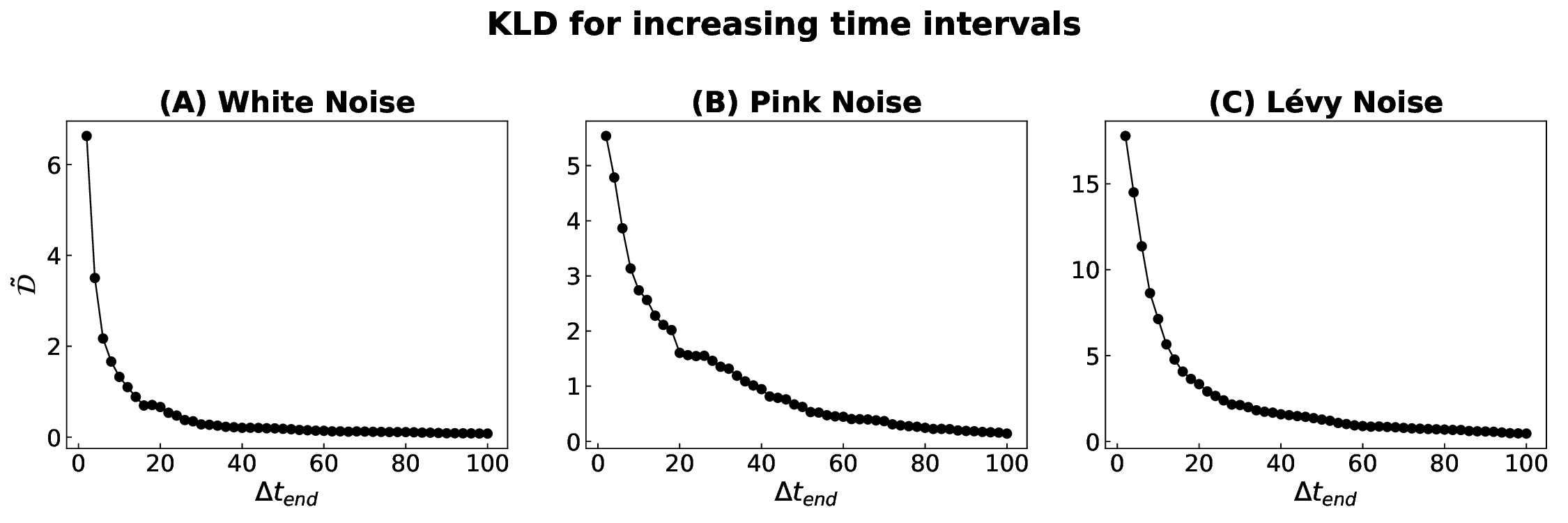}
\caption{\label{Fig2}KLD for increasing time intervals, showing the approach to zero for (A) Gaussian white noise, (B) Pink noise, (C) Lévy noise. This means that the lower bound of EP saturates to zero at larger time scales, which implies the equilibrium or steady state of the system at macroscopic scale.}
\end{center}
\end{figure*}

As illustrated before, the scheme of the increasing time intervals began with the interval 0-2 and then progressed by increasing the time window length by 2 arbitrary units. This will provide the cumulative behavior of the system's dynamics. The average KLD, $\tilde{\mathcal{D}}$ at each interval is plotted against the end points of each time interval, $\Delta t_{\text{\textit{end}}}$ [Fig.\ \eqref{Fig2}]. For all forms of noise, $\tilde{\mathcal{D}}$ starts from a non-zero value and approaches zero at larger time scales. This shows that the lower bound of EP saturates to zero at larger time scales. So, macroscopically, the local dynamics governed by fluctuations has been averaged out to give an overall cumulative behavior that implies the equilibrium or steady state of the system.

\begin{figure*}
\begin{center}
\includegraphics[height=5.0cm,width=15.0cm]{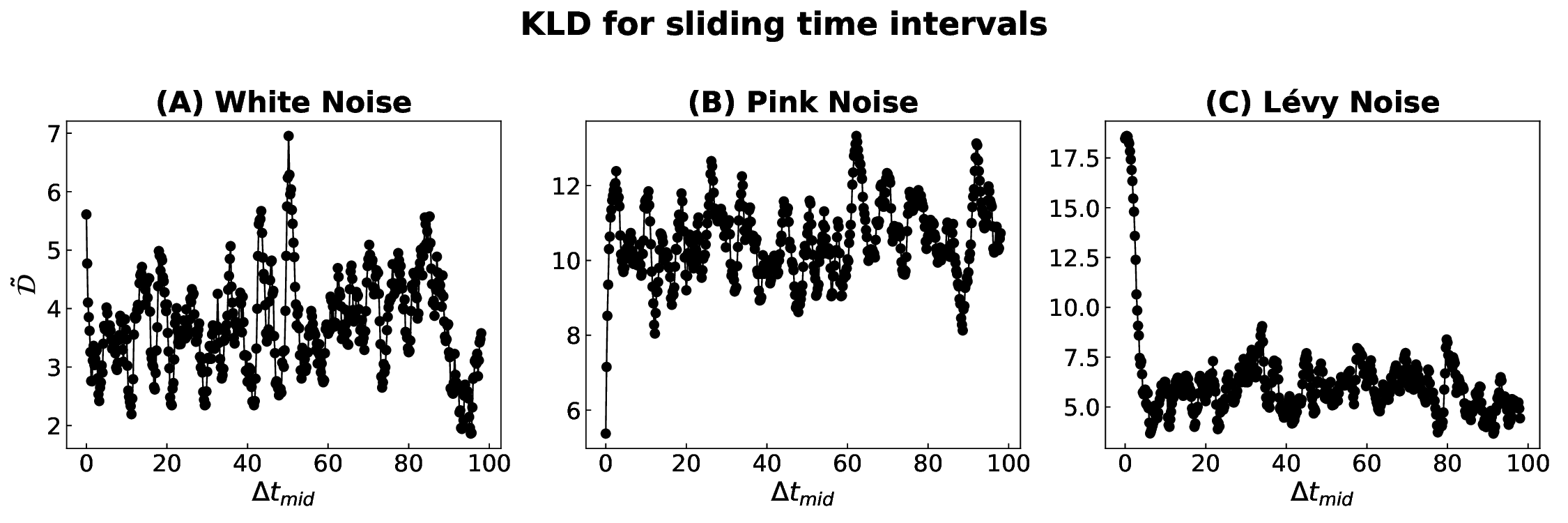}
\caption{\label{Fig3}KLD for sliding time intervals, showing random values for (A) Gaussian white noise, (B) Pink noise, (C) Lévy noise. This means that the lower bound of EP does not approach zero and hence the nonequilibrium dynamics is prominent at the shorter time scales. For pink noise, $\tilde{\mathcal{D}}$ increases initially from a small value and for Lévy noise, $\tilde{\mathcal{D}}$ decreases initially from a larger value.}
\end{center}
\end{figure*}

For the sliding time intervals with a fixed time window length of 2 arbitrary units, we plotted $\tilde{\mathcal{D}}$ against the mid points of each time interval, $\Delta t_{\text{\textit{mid}}}$ [Fig.\ \eqref{Fig3}]. The values of $\tilde{\mathcal{D}}$ are random which indicates that the observations at the shorter time scales reveal the inherent fluctuations in the system that has been suppressed in the macroscopic scale. Subsequently, this implies that the lower bound of EP does not approach zero and hence the nonequilibrium dynamics is prominent at the shorter time scales. The time reversal asymmetry that we observed from the trajectories itself has been reflected significantly through these observations.

Apart from the qualitative disordered behavior depicted from the observations at the shorter time scales, it also gives the difference in magnitudes of the values of  $\tilde{\mathcal{D}}$ for the different forms of noise. We find that $\tilde{\mathcal{D}}$ increases initially from a smaller value at the beginning of the simulation time for pink noise [Fig.\ \eqref{Fig3}B]. As this result is obtained after multiple realisations, we can induce that at the beginning of the simulation time, the forward and backward trajectories overlap. Then, as time progresses, the trajectories diverge due to the long range temporal correlation of the noise. On the other hand, $\tilde{\mathcal{D}}$ decreases initially from a larger value at the beginning of the simulation time for Lévy noise [Fig.\ \eqref{Fig3}C], which suggests that there is an incremental jump at one of the trajectories that produces such a large value of $\tilde{\mathcal{D}}$. All of these results are obtained for the same time window length of 2 arbitrary units.

\subsection{Dependence on time window length}
\begin{figure*}
\begin{center}
\includegraphics[height=7.0cm,width=8.0cm]{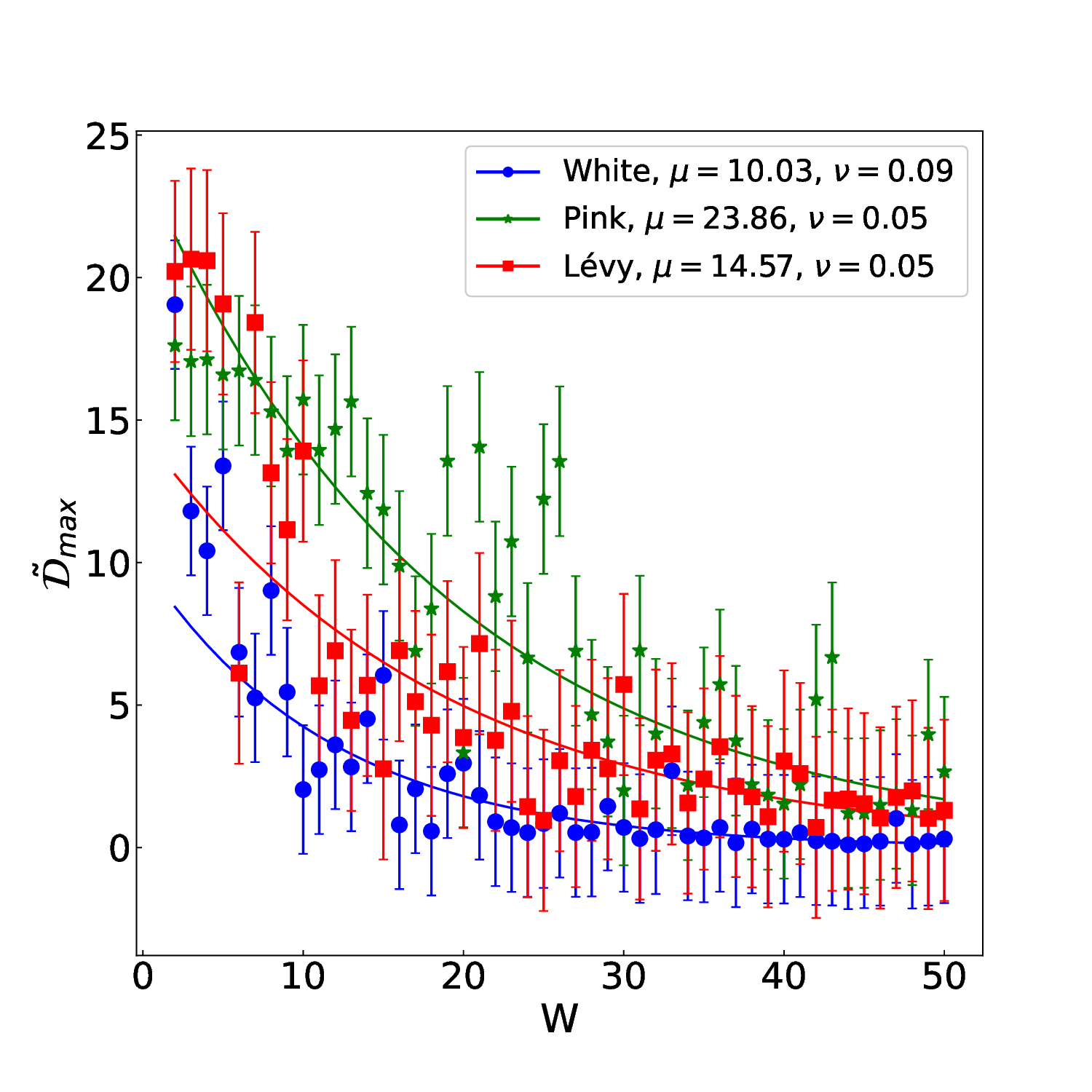}
\caption{\label{Fig4}Maximum average KLD values ($\tilde{\mathcal{D}}_{\text{max}}$) with respect to the length of the sliding time window, W.}
\end{center}
\end{figure*}

To check the effect of using larger sliding time window lengths, we determined the maximum value of the average KLD value, $\tilde{\mathcal{D}}_{\text{max}}$ at each time window for every form of noise. We found that $\tilde{\mathcal{D}}_{\text{max}}$ decreases exponentially for every form of noise [Fig.\ \eqref{Fig4}]. This is understandable since we have already observed that the KLD values decrease as we increase the time intervals [Fig.\ \eqref{Fig2}]. Thus, we can formulate a relation for $\tilde{\mathcal{D}}_{\text{max}}$ as a function of the time window length, $W$, as
\begin{eqnarray}
\label{D_and_W}
\tilde{\mathcal{D}}_{\text{max}}(W)=\mu\exp(-\nu W)
\end{eqnarray}
where $\mu$ and $\nu$ are positive constants that depend on the form of noise. They are determined according to the exponential fits to the $\tilde{\mathcal{D}}_{\text{max}}$ values. $\nu$ can be called as the decay rate of $\tilde{\mathcal{D}}_{\text{max}}$ with respect to $W$.

We can see from Fig.\ \eqref{Fig4} that the magnitude of $\tilde{\mathcal{D}}_{\text{max}}$ is largest for pink noise and is the least for white noise at each time window. This can also be explained by the long range temporal correlation feature of the pink noise distribution and the uncorrelated nature of white noise distribution. On the other hand, for Lévy noise, the magnitude of $\tilde{\mathcal{D}}_{\text{max}}$ lies midway due to its random increments that define its intermediate characteristics.

We also determined the minimum time window length at which $\tilde{\mathcal{D}}_{\text{max}}$ equilibrates to a constant value. This is done by calculating the differences between the $\tilde{\mathcal{D}}_{\text{max}}$ values at successive time windows from the exponential fit such that the difference is just less than 0.1. For white noise, it is found to be at $W_{\text{eq}}^{(\text{White})}=26$, for Lévy noise, $W_{\text{eq}}^{(\text{Lévy})}=39$ and for pink noise, $W_{\text{eq}}^{(\text{Pink})}=49$. We also kept the maximum time window length for observation at 50, so that the time window length is always less than or equal to half the total simulation time. 

\begin{figure*}
\begin{center}
\includegraphics[height=6.0cm,width=12.0cm]{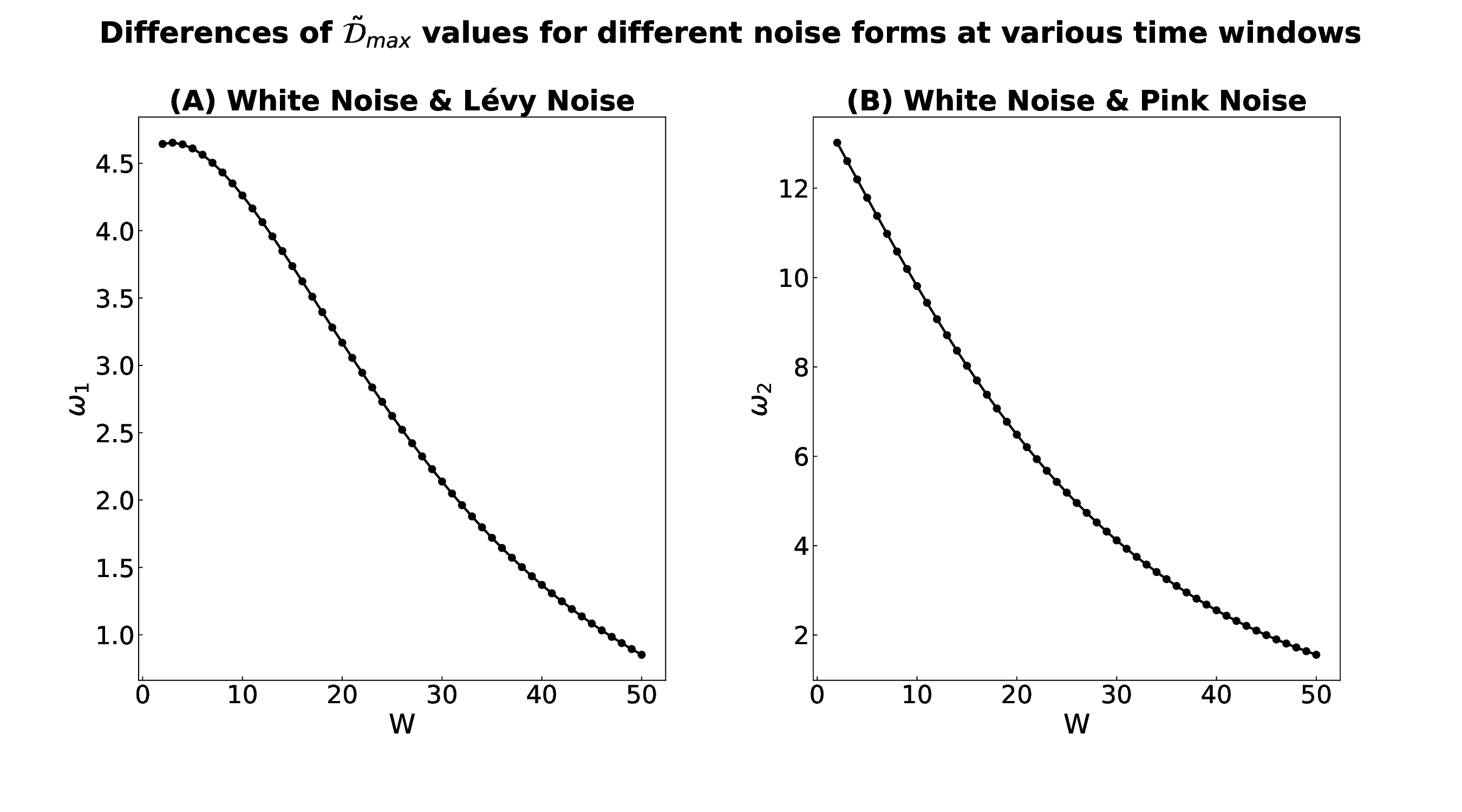}
\caption{\label{Fig5}Differences of $\tilde{\mathcal{D}}_{\text{max}}$ values [$\omega_1=\tilde{\mathcal{D}}_{\text{max}}^{(\text{Lévy})}-\tilde{\mathcal{D}}_{\text{max}}^{(\text{White})}$ and $\omega_2=\tilde{\mathcal{D}}_{\text{max}}^{(\text{Pink})}-\tilde{\mathcal{D}}_{\text{max}}^{(\text{White})}$] with respect to the length of the sliding time window, W.}
\end{center}
\end{figure*}

Finally, we determined the differences between the values of $\tilde{\mathcal{D}}_{\text{max}}$ for the different forms of noise at each time window. We set $\omega_1=\tilde{\mathcal{D}}_{\text{max}}^{(\text{Lévy})}-\tilde{\mathcal{D}}_{\text{max}}^{(\text{White})}$ and $\omega_2=\tilde{\mathcal{D}}_{\text{max}}^{(\text{Pink})}-\tilde{\mathcal{D}}_{\text{max}}^{(\text{White})}$ and observed that they decrease as $W$ increases [Fig.\ \eqref{Fig5}]. This shows that observations at larger time scales show no distinction between the various noise forms, which is again consistent with the case of the increasing time intervals.\\

\section{Conclusion}
The measurement of the lower bound of EP provides valuable insights into the dynamical behavior of stochastic systems. A clear distinction exists between the observations at the global macroscopic scale and the local dynamics at short time scales. By determining the KLD over sliding time intervals, we explored the small-scale local dynamics at short time scales for different forms of noise accompanying the particle's trajectory. This approach highlights the disordered yet intricate nature of microscopic dynamics that is often overshadowed by the steady-state behavior observed on larger scales.

Although the macroscopic time-scale observations depict a zero lower bound for EP - indicating a nonequilibrium steady state - the time-reversal asymmetry and the breaking of detailed balance become apparent only at the short time scales. This result suggests that the ordered behavior observed at the macroscopic scale may emerge from small-scale disorders at the microscopic scale, a phenomenon typically characteristic of fractal behavior \cite{Schroeder}. Understanding this interplay between scales can provide deeper insights into how systems self-organize and maintain nonequilibrium steady states. Future studies could explore the presence of fractal structures by quantifying specific features, such as scaling laws or self-similar patterns, which define the system's overall dynamical nature. Such analyses might also reveal connections to entropy generation and dissipation in broader nonequilibrium contexts.

Examining different forms of noise reveals how various fluctuations manifest distinct behaviors within the same stochastic system. Gaussian white noise, with its lack of temporal correlation, serves as a baseline for understanding randomness in simpler systems. In contrast, pink noise and Lévy noise, which are prevalent in real-world complex systems, demonstrated distinct impacts on time-reversal asymmetry and entropy production. Pink noise, with its long-range correlations, exhibits stronger deviations at smaller scales, while Lévy noise, characterized by heavy-tailed distributions and sudden jumps, reflects intermediate behavior between white and pink noise. The need for making observations and studying systems at short time scales is highlighted in the fact that on the macroscopic scale, these noise forms show no distinction.

Our findings also demonstrate how the techniques employed successfully capture the inherent properties of the diverse noise forms, underscoring their utility in modeling complex stochastic dynamics. This study could be extended to nonlinear or coupled stochastic differential equations (SDEs) \cite{Rus}, facilitating the development of novel methods for investigating disorder in nonequilibrium systems. Additionally, multi-variable and multi-dimensional systems could be analyzed to determine how the current techniques might be refined to yield more robust results. \\

\begin{table*}
\begin{center}
\label{Table I}
\begin{tabular}{|c | c|} 
 \hline
 Symbols & Description \\ [0.5ex] 
 \hline\hline
 $x$ & position of the particle \\
 \hline
 $t$ & time \\ 
 \hline
 $a(x,t)$ & drift term \\
 \hline
 $b(x,t)$ & diffusion diffusion \\
 \hline
 $\zeta$ & noise or fluctuation term \\
 \hline
 $n$ & order of numerical iteration \\ 
 \hline
 $\Delta\phi$ & $\zeta\Delta t$ \\
 \hline
 $\mathcal{D}$ & Kullback-Leibler divergence (KLD) \\
 \hline
 $\rho$ & probability density function of forward trajectory \\
 \hline
 $\bar{\rho}$ & probability density function of backward trajectory \\
 \hline
 $\langle\Delta S_{\text{tot}}\rangle$ & average entropy production of the whole system \\
 \hline
 $k$ & Boltzmann constant, set to unity \\
 \hline
 $W$ & time window length \\
 \hline
 $\sigma$ & standard deviation of noise distribution \\
 \hline
 $\alpha$ & stability parameter of Lévy noise \\
 \hline
 $\beta$ & skewness of Lévy noise \\
 \hline
 $\epsilon$ & stability parameter added to the probability density value \\
 \hline
 $\tilde{\mathcal{D}}$ & average KLD value \\
 \hline
 $\Delta t_{\text{end}}$ & end time of increasing time intervals \\
 \hline
 $\Delta t_{\text{mid}}$ & mid point of sliding time intervals \\
 \hline
 $\tilde{\mathcal{D}}_{\text{max}}$ &  maximum of average KLD values \\
 \hline
 $\mu$ &  positive constant \\
 \hline
 $\nu$ &  positive constant, decay rate of maximum of average KLD \\
 & values with time window length \\
 \hline
 $W_{\text{eq}}^{(\text{White})}$, $W_{\text{eq}}^{(\text{Lévy})}$ and $W_{\text{eq}}^{(\text{Pink})}$ & minimum time window length at which $\tilde{\mathcal{D}}_{\text{max}}$ equilibrates \\
 \hline
 $\omega_1$ &  $\tilde{\mathcal{D}}_{\text{max}}^{(\text{Lévy})}-\tilde{\mathcal{D}}_{\text{max}}^{(\text{White})}$ \\
 \hline 
 $\omega_2$ &  $\tilde{\mathcal{D}}_{\text{max}}^{(\text{Pink})}-\tilde{\mathcal{D}}_{\text{max}}^{(\text{White})}$ \\ [1ex]
 \hline
\end{tabular}
\caption{\label{Table I}Symbols and notations used throughout the article.}
\end{center}
\end{table*}

{\noindent}\textbf{Acknowledgement:}
MKS is a Junior Research Fellow (JRF) under the National Fellowship for Scheduled Castes Students (NFSC) scheme and acknowledges the National Scheduled Castes Finance and Development Corporation (NSFDC) and the Ministry of Social Justice \& Empowerment, Government of India for providing financial support. The authors would like to thank Dr. Athokpam Langlen Chanu, Preet Mishra and Shyam Kumar for valuable discussions.\\
\textbf{Author Contributions:} MKS and RKBS conceptualised the work. MKS and MSS did the analytical and computational work. All authors have accepted responsibility for the entire content of this manuscript and approved its submission.\\
\textbf{Competing interests:} The authors state no conflict of interest.\\
\textbf{Data Availability:} Not applicable.\\

\pagebreak

\end{document}